\documentclass{article}

\usepackage{emulateapj}
\usepackage{onecolfloat}
\usepackage{graphicx} 
\usepackage{fancyheadings} 
\usepackage{ulem}
\usepackage{rotating}
\usepackage{lscape}

\newcommand{\tskip}{\tablevspace{3pt}}
\newcommand{\citep}{\cite}
\newcommand{\feh}{[{\rm Fe/H}]}

\newcommand{\lya}{Ly$\alpha$ }

\newcommand{\cm}[1]{\, {\rm cm^{#1}}}
\newcommand{\N}[1]{{N({\rm #1})}}

\newcommand{\rAA}{{\AA \enskip}}

\newcommand{\smm}{\sum\limits}

\begin{document}

\twocolumn[%
\accepted{ApJ: February 28, 2000}

\title{METALLICITY EVOLUTION IN THE EARLY UNIVERSE}

\author{ JASON X. PROCHASKA\altaffilmark{1} \\
The Observatories of the Carnegie Institute of Washington
813 Santa Barbara St. \\
Pasadena, CA 91101 \\
and \\
ARTHUR M. WOLFE\altaffilmark{1} \\
Department of Physics, and Center for Astrophysics and Space Sciences \\
University of California, San Diego \\
C--0424; La Jolla; CA 92093}

\begin{abstract} 

Observations of the damped \lya systems provide
direct measurements on the chemical enrichment history of neutral gas in the 
early universe.  In this Letter, we present new measurements for four damped
\lya systems at high redshift.  Combining these data
with $\lbrack$Fe/H$\rbrack$ values culled from the literature,  we investigate
the metallicity evolution of the universe from $z \approx 1.5-4.5$.
Contrary to our expectations and the predictions of essentially every 
chemical evolution model, the $\N{HI}$-weighted mean $\lbrack$Fe/H$\rbrack$ 
metallicity exhibits minimal evolution over this epoch.
For the individual systems,
we report tentative evidence for an evolution in 
the unweighted $\lbrack$Fe/H$\rbrack$
mean and the scatter in $\lbrack$Fe/H$\rbrack$ with the higher 
redshift systems showing
lower scatter and lower typical $\lbrack$Fe/H$\rbrack$ values.  
We also note that no damped \lya system has $\lbrack$Fe/H$\rbrack$ 
$< -2.7$~dex. Finally, we discuss the potential
impact of small number statistics and dust on our conclusions and 
consider the implications of these results
on chemical evolution in the early universe.

\end{abstract}

]

\altaffiltext{1}{Visiting Astronomer, W.M. Keck Telescope.
The Keck Observatory is a joint facility of the University
of California and the California Institute of Technology.}

\pagestyle{fancyplain}
\lhead[\fancyplain{}{\thepage}]{\fancyplain{}{PROCHASKA \& WOLFE}}
\rhead[\fancyplain{}{METALLICITY EVOLUTION IN THE EARLY UNIVERSE}]
{\fancyplain{}{\thepage}}
\setlength{\headrulewidth=0pt}
\cfoot{}

\section{INTRODUCTION}

The damped \lya systems, neutral hydrogen gas layers identified in the
absorption line spectra of background quasars, 
dominate the neutral hydrogen 
content of the Universe at all epochs.   At high redshift,
these systems are widely
accepted as the progenitors of present-day galaxies for the following
reasons:
(i) their very large HI
column densities, $\N{HI} > N_{thresh} = 2 \times 10^{20} \cm{-2}$, imply
overdensities $\delta \rho / \rho \gg 100$, i.e., these are virialized
systems at high redshift; 
(ii)  they contain the majority of neutral gas 
in the early Universe and are therefore the reservoirs for galaxy
formation; (iii) their gas density
$\Omega_{gas}$ at redshift $z \approx 2 -3$ is consistent with the
mass density of stars today (\citep{wol95}).  
While the physical nature
of the damped \lya systems is still controversial 
(\citep{pro97,hae98,maller99,lebr97}), 
by studying the chemical abundances of the damped \lya system one directly
traces the chemical enrichment history of the Universe at high redshift.
Observing damped \lya systems is equivalent to
poking sightlines through the ISM of protogalaxies. 
Because these observations are biased to HI cross-section
and the HI gas mass of a system is proportional to
$\int \sigma \cdot N$,
one can measure global properties of the universe 
simply by weighting the measurement from 
each damped system by $\N{HI}$. 
At the same time,  the observations afford an efficient means for
examining the characteristics of individual protogalaxies
in the early Universe.
In this Letter, we examine the metallicity of
the damped \lya systems from $z \approx 2 - 4.5$ which places
tight constraints on chemical evolution models (e.g.\ Pei et al.\ 1999),
as well as a valuable consistency check on SFR observations
(\citep{ptt99a}).

Over the past decade, several groups have surveyed the metallicity of the
damped \lya systems from $z \approx 1 - 4$ 
(\citep{ptt94,ptt97,ptt99a,lu96,pro99}). 
To date, the chemical abundances of over 40 systems have been measured, 
the majority with $z = 1.5 - 3$
where the identification and follow-up observations
of damped \lya systems is most efficient.
These studies argue that at $z \approx 2$, the mean metallicity
of the damped systems is approximately 1/10 $-$ 1/30 solar metallicity
([Zn/H]~$\approx -1.1$,
[Fe/H] $\approx -1.5$) with a large scatter from nearly solar to less than
$1/100$ solar metallicity.  At very high redshift ($z>3$), the picture
is far less certain.  Focusing on a sample of seven $z>3$ damped \lya
systems, Lu et al.\ (1996,1997) noted
that the metallicity of these systems is significantly
lower than the $z<3$ observations.  
In turn, the authors argued that $z \approx 3$ marked the epoch 
where significant star formation begins,
a claim with important implications for the processes of galaxy formation.

In this Letter we present new measurements on the metallicity of
four damped \lya systems (including three at $z>3.5$) and together 
with the data from Prochaska \& Wolfe (1999) double the sample of $z>3$
systems. The new full sample -- including the systems from 
Lu et al.\ (1996,1997) --
reveals evidence for little change in the $\N{HI}$-weighted
mean metallicity of the neutral Universe from 
$z \approx 1.5-4.5$, contrary to the predictions of essentially every chemical
evolution model.   On the other hand, we find tentative evidence for an
evolution in the unweighted mean and scatter of [Fe/H] for individual damped 
\lya systems. Finally, we comment on the
robustness of these results (particularly in the light of small number
statistics and dust), speculate on the implications
for chemical enrichment, and discuss the prospects for future advances.

\section{OBSERVATIONS AND ANALYSIS}
\label{sec-obs}

To determine the metallicity of a damped \lya system, one must accurately
measure the neutral hydrogen column density $\N{HI}$ and a metallicity
indicator,  typically either Zn or Fe.
In stellar population studies of the Galaxy one traditionally uses Fe
as the metallicity indicator, primarily as a matter of convenience.  
As we are
studying gas-phase abundances, however, we must account for the
possible depletion of Fe onto dust grains or instead choose an
element like Zn which is minimally affected by depletion.  Unfortunately,
there are both
theoretical and observational disadvantages to using Zn as the
metallicity indicator.  Theoretically,
Zn has a very uncertain chemical origin.
It is referred to as an Fe peak element because it traces
Fe in Galactic stars (\citep{sne91}),  yet the leading theory on the production
of Zn proposes it forms in the neutrino-driven 
winds of Type~II SN (\citep{hff96}).  
Furthermore, recent measurements of [Zn/Fe] in
metal-poor stars (\citep{jhnsn99}) and thick disk stars
(\citep{pro00}) suggest Zn/Fe is enhanced relative to the Sun by
+0.1 to +0.3~dex, perhaps consistent with a Type~II origin.
Observationally there are complications with measuring Zn in the 
damped \lya systems, where one must rely on two
weak ZnII transitions with $\lambda_{rest} \approx 2000$\AA.
The transitions are so weak that 
even at high resolution and high S/N, Zn can only be detected in damped
systems when $\log [\N{HI}] + [{\rm Zn/H}] > 19.0$ (e.g.\ [Zn/H]~$> -1.3$ for
systems with $\N{HI} \approx N_{thresh}$).  Most important to this study,
however, the large rest wavelength of the ZnII transitions prevents one
from readily measuring Zn in $z>3$ damped \lya systems as it is 
difficult to make sensitive observations at $\lambda \approx 8000$\rAA
with current high resolution spectrographs.  
In fact, at the time of publication {\it we are not aware of a single accurate
Zn measurement for any $z>3$ damped \lya system}.
Therefore, we will focus
on Fe in this Letter, which has two singly ionized transitions at 
$\lambda_{rest} \approx 1600$\rAA with a complement of $f$-values
ideal for measuring the abundance of Fe in 
damped systems.  We restrict the analysis
to Fe measurements made with HIRES on the Keck~I telescope (\citep{vgt92}), 
specifically the systems
observed by Prochaska \& Wolfe (1999) and Lu et al.\ (1996,1997) 
and the additional systems
introduced here.  In addition to providing a homogeneous data set which has
been reduced and analyzed with the same techniques,
these observations account for nearly every damped \lya system
with an accurate Fe abundance at $z>1.5$ and every system with $z>3$.

\begin{table}[ht] \footnotesize
\begin{center}
\caption{ \label{newobs}}
{\sc NEW METALLICITY MEASUREMENTS} 
\begin{tabular}{llccccc}
\tskip
\tableline
\tableline \tskip
QSO & $z_{abs}$& $\N{HI}$    & $\N{Fe^+}$ & [Fe/H] \\
\tableline \tskip
BRI0952$-$0115 & 4.024 & $20.50 \pm 0.1$ & $14.054 \pm 0.07$\tablenotemark{a} 
& $-1.95$ \\
BRI1108$-$0747 & 3.608 & $20.50 \pm 0.1$ & $13.860 \pm 0.03$\tablenotemark{b} 
& $-2.14$ \\
Q1223$+$1753   & 2.466 & $21.50 \pm 0.1$ & $15.279 \pm 0.03$\tablenotemark{c} 
& $-1.72$ \\
PSS1443+2724   & 4.224 & $20.80 \pm 0.1$ & $15.325 \pm 0.10$\tablenotemark{c} 
& $-0.98$ \\
\tskip \tableline
\end{tabular}
\end{center}
\centerline{$^a$Average of FeII 1144 ($\log gf = 0.105$) and FeII 1608}
\centerline{$^b$FeII 1608}
\centerline{$^c$FeII 1611}
\end{table}

Table~\ref{newobs} summarizes the new $\N{Fe^+}$ measurements derived from
observations acquired by the authors in  February 1998 and March 1999 
with HIRES on the Keck~I 10m telescope.  The data was reduced with the
{\it makee} software package developed by T. Barlow
and the column densities were 
derived primarily from the FeII~1608 and/or FeII~1611 
transitions with the apparent optical
depth method (\citep{sav91}).  We adopt the oscillator strengths from 
Cardelli \& Savage (1995) 
noting that our conclusions on the evolution of [Fe/H] in
the damped \lya systems are not sensitive to their values.
The $\N{HI}$ values for these systems are taken from the literature
(\citep{wol95,storr99}) and are the dominant source of error in the 
[Fe/H] values.
Finally, we evaluate [Fe/H] assuming the meteoritic Fe abundance
($\epsilon$(Fe)=7.50; Grevesse \& Sauval 1999)
without adopting any ionization corrections
which is an excellent assumption for all but possibly the lowest 
$\N{HI}$ damped \lya systems (\citep{pro96}).
Together with the published measurements of Prochaska \& Wolfe (1999) and 
Lu et al.\ (1996,1997) the total [Fe/H] sample is 37 systems, 15 with 
$z>3$.  The systems were chosen independent of any prior metallicity
measurements; the only possible biases are due to the magnitude limited
selection of the quasars (e.g.\ Fall \& Pei 1993)
which will be discussed in the following section. 

\section{RESULTS AND DISCUSSION}
\label{sec-disc}

\begin{figure*}[ht]
\begin{center}
\includegraphics[height=5.3in, width=3.7in,angle=-90]{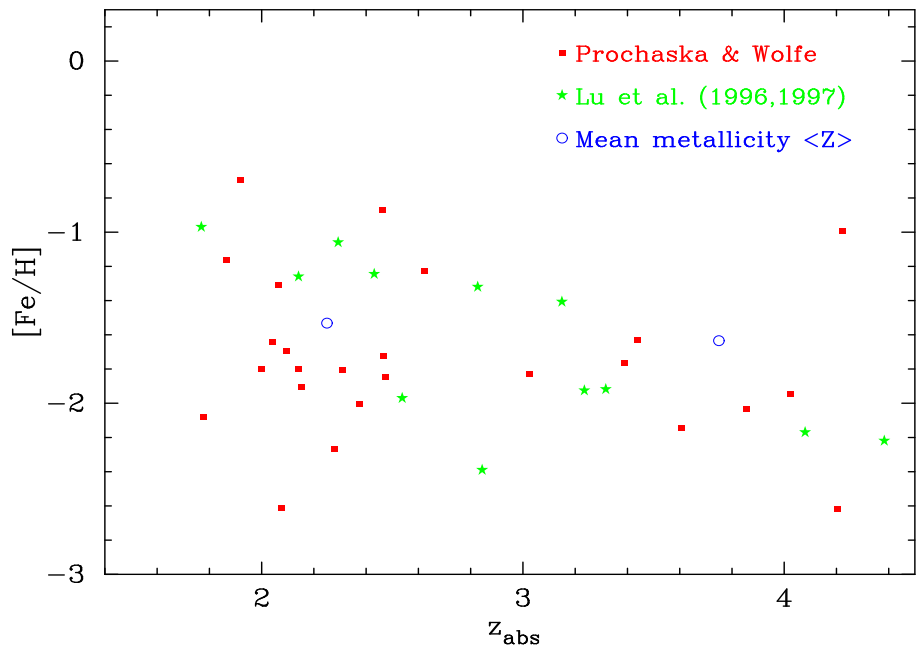}
\caption{Thirty-nine [Fe/H], $z_{abs}$ pairs for the damped \lya systems
observed by Prochaska \& Wolfe (1999, this paper; squares) 
and Lu et al.\ (1996,1997; 
stars) with HIRES on the Keck~I telescope.  The open circles correspond to
the $\N{HI}$-weighted mean metallicity for the systems at 
$z_{abs} = [1.5,3]$ and $z_{abs} = (3,4.5]$.  Note that the difference in these
means is small, $\approx 0.1$~dex.  Also observe that the scatter in the [Fe/H]
values appears to increase at lower redshift.}
\label{Fevsz}
\end{center}
\end{figure*}

Figure~\ref{Fevsz} plots the 37 [Fe/H] values versus $z_{abs}$ for 
the Wolfe \& Prochaska
sample (dark squares) and (light stars)
the sample of damped \lya systems observed by
Lu et al.\ (1996,1997).  To explore evolution in the metallicity of the damped
\lya systems,
we consider three moments of the metallicity data 
in two redshift intervals, $z_{low} = [1.5,3]$ and
$z_{high} = (3,4.5]$.  These are: 
(1) the $\N{HI}$-weighted mean metallicity of neutral gas, $<Z>$;
(2) the unweighted mean metallicity, $<\feh>$,
of the set of damped \lya systems at $z_{low}$ and $z_{high}$; and 
(3) the standard deviation of [Fe/H] in these protogalaxies, 
$\sigma(\feh)$.  

The first moment represents the global metallicity
of all neutral gas at a given epoch, $\Omega_{metals}/\Omega_{HI}$.  
It is evaluated by weighting each
[Fe/H] measurement by the corresponding HI column density, 
$<Z> \; \equiv \, \log [\smm \N{Fe^+} / \smm \N{HI}] - 
\log ( {\rm Fe/H} )_\odot$.
Computing
the mean for the damped \lya systems at the two
epochs, we find $<Z>_{low} = -1.532 \pm 0.036$ and 
$<Z>_{high} = -1.634 \pm 0.049$. 
The errors on the $<Z>$ values reflect only
the statistical uncertainty in measuring
$\N{Fe^+}$ and $\N{HI}$ and
were derived with standard error propagation techniques. 
Below we estimate the uncertainty due to small number statistics.
Comparing the $<Z>$ values, we note that they
favor no significant evolution
in the mean metallicity of neutral gas from $z = 1.5 - 4.5$.
If we include the tentative result from Pettini et al.\ (2000)
that the Zn mean metallicity does not change from $z \approx 1 - 3$, then
one concludes there is no evidence for
significant metallicity evolution from $z=1-4.5$,
an interval spanning more than 3~Gyr.  The other two
moments, the unweighted mean
$<\feh> = \frac{1}{n} \smm^n \feh$ and the scatter $\sigma(\feh)$, 
are more sensitive to the chemical enrichment history within 
individual protogalaxies as each damped system is given equal weight.
For the two intervals we find that the mean logarithmic
abundance, $<\feh>_{low} = -1.61$ and $<\feh>_{high} = -1.83$.
Meanwhile, the scatter in [Fe/H] is $\sigma(\feh) = 0.50$ and
$\sigma(\feh) = 0.35$ for the $z_{low}$ and $z_{high}$ samples
respectively.   Performing the Student's t-test and the F-test 
on the two moments, we find that the
$<\feh>$ and $\sigma(\feh)$ statistics for the two epochs
are inconsistent at the 90$\%$
and 80$\%$ c.l.  Therefore, there is tentative evidence for chemical 
evolution in the individual damped \lya systems with the $z<3$ sample
showing a higher typical metallicity and a larger scatter in [Fe/H]
from system to system.

\begin{figure*}[ht]
\begin{center}
\includegraphics[height=5.3in, width=3.7in,angle=-90]{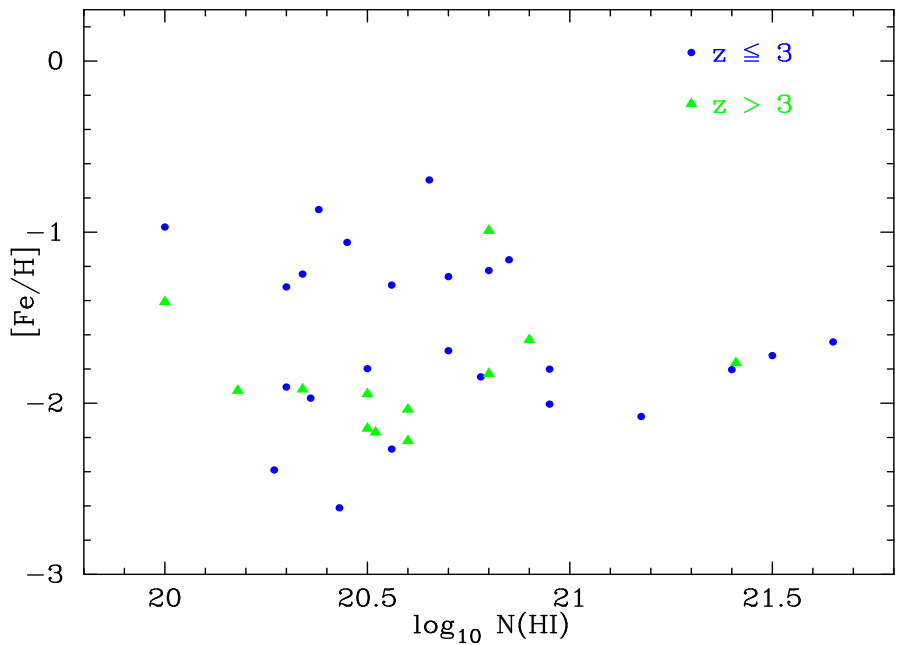}
\caption{Thirty-nine [Fe/H], $\N{HI}$ pairs for the damped \lya systems
in the full sample.  The circles correspond to $z_{abs} \leq 3$ systems and
the triangles are for $z_{abs} > 3$.  While the systems with 
$\N{HI} < 10^{21} \cm{-2}$ show a large scatter in [Fe/H], the large
$\N{HI}$ systems all have $\feh \approx -1.8$.}
\label{FeHvsNHI}
\end{center}
\end{figure*}

To address the robustness of these results, one must consider several issues.
First, because $<Z>$ is dominated by the systems with the largest
$\N{HI}$ values, this mean is robust only in so far as the total $\N{HI}$,
HI$_T$~$\equiv \smm_n \N{HI}$, 
well exceeds that of a single damped \lya system.  
Figure~\ref{FeHvsNHI} plots the $\feh$, $\N{HI}$ pairs for all 37 
systems where the dark circles are members of the $z_{low}$ sample and the
light triangles those of the $z_{high}$ sample.
Note that there are four systems with $\N{HI}>10^{21} \cm{-2}$ at
$z<3$ but only a single system in the $z_{high}$ sample. For the
$z_{low}$ sample, HI$_T = 10^{22.30} \cm{-2}$, which is a factor of
4 larger than the largest $\N{HI}$ measured for any damped \lya system
(Q0458$-$0203; $\N{HI} = 10^{21.7}$) and 10 times greater than 
most of the known damped \lya systems.  As such, we consider the mean
derived from the $z_{low}$ sample to be reasonably robust.  The primary
potential
pitfall is if the optical surveys have systematically missed damped systems
with $\N{HI} > 10^{22} \cm{-2}$, a possibility if dust obscuration is
significant (discussed further below).
The situation is far more uncertain for the $z>3$ sample where 
HI$_T = 10^{21.84} \cm{-2}$, comparable to the $\N{HI}$ of
the Q0458$-$0203 system from the $z_{low}$ sample
and only three times greater than the largest $\N{HI}$ system in the
$z_{high}$ sample.  While the most recent surveys 
suggest there are very few $z>3$ damped systems with 
$\N{HI} > 10^{21} \cm{-2}$ Storrie-Lombardi \& Wolfe (2000), 
we caution that the mean
we have derived for the $z_{high}$ sample is a tentative result.  
For example, the system toward Q0000$-$2619 has significant bearing on 
$<Z>_{high}$ and its Fe abundance has been difficult to determine
(\citep{pro99,lu96}).  Ironically, removing it from the $z_{high}$
sample would actually increase $<Z>_{high}$ into exact
agreement with $<Z>_{low}$ because we have adopted 
[Fe/H]~$= -1.77$ based on the FeII~1611 profile from this system.  
Meanwhile, lowering [Fe/H] 
by 0.6~dex to establish consistency with the Ni and Cr abundances, 
would decrease $<Z>_{high}$ by 0.1~dex. 
In short, while we have confidence in the $<Z>_{low}$ value,
we caution the reader that 
small number statistics are still important
in evaluating the $<Z>_{high}$.
One can estimate the uncertainty associated with the small number statistics
of the two samples by performing a bootstrap statistical analysis.
For each sample, we independently
calculated $<Z>$ 500 times by randomly drawing 
$n$ objects ($n$ is the number of damped systems in a given redshift 
interval) from each interval.  In turn, we can estimate the effects of cosmic
variance on our results by calculating the standard deviation of the
two bootstrap $<Z>$ distributions: $\sigma^{<Z>}_{low} = 0.088$~dex and
$\sigma^{<Z>}_{high} = 0.155$~dex.
As one would expect, the $<Z>_{high}$ value, which is based on 
only 15 systems, is 
considerably less certain than the $<Z>_{low}$ measurement.
The difference in the $\sigma^{<Z>}$ values
stresses the outstanding need for
future observational programs to focus on $z>3$ damped systems.

Any study on the chemical abundances of the damped \lya systems must
assess the potential effects of dust.  
With respect to this analysis, where
we have taken Fe as the metallicity indicator, there are two important
aspects to consider: (1) if we need to correct the
observed [Fe/H] values by some factor to obtain the true metallicity of
each system, does the mean correction evolve in
time and/or differ from system to system at the same epoch?; 
and (2) dust obscuration could remove 
damped \lya systems from the magnitude limited 
samples which would significantly alter the conclusions (e.g.\ metal-rich,
high $\N{HI}$ systems).  With respect to the first concern, we can estimate
the maximum dust correction to [Fe/H] via the measured Zn/Fe ratio.
Again, Zn is essentially undepleted in the gas-phase so that
[Zn/H] = [Fe/H] + [Zn/Fe] may be more representative of the true metallicity
in the damped \lya systems.
This practice is limited, however, by the fact that Zn may be produced
in Type~II SN (\citep{hff96}) such that super-solar Zn/Fe ratios would be 
representative of nucleosynthesis, not dust depletion.
Furthermore, recent results on the [Zn/Fe] ratio measured in Galactic stars 
shows that [Zn/Fe] $> +0.2$~dex in very metal-poor stars ($\feh < -2.5$;
Johnson 1999) and even exhibits super-solar values 
(average [Zn/Fe]~$\approx +0.13$ in 10 stars; Prochaska et al.\ 2000) 
in thick disk stars with $\feh > -1$. 
Therefore, while the typical [Zn/Fe] value in the damped \lya systems is
+0.4~dex with relatively small scatter (\citep{ptt97,pro99}), 
it is unclear what fraction 
is due to dust depletion.  Nonetheless, if we take [Zn/H] as the
true metallicity indicator, $<Z>$ and the
unweighted mean are enhanced 
by $\approx 0.4$~dex, but there is
very little change in the observed scatter.
The potential effects of dust depletion on 
the statistical moments for the $z_{high}$ sample are more
speculative as there is {\it no accurate Zn determination} for any
$z>3$ damped \lya system.  To estimate the depletion level, we can compare
the relative abundance patterns 
(in particular the Si/Fe ratio) of these systems with the
$z_{low}$ sample. 
In the few cases where Si/Fe has been measured in the $z>3$ systems
one finds [Si/Fe]~$\approx +0.3$~dex, nearly identical to
the typical value of the $z<3$ sample.
While the similarity of a metal ratio like Si/Fe does not require 
similar dust depletion levels, the $z>3$ [Si/Fe] values do
imply depletion levels of at least 0.3~dex. 
Therefore, unless one takes the unlikely stance that
the $z>3$ systems are significantly more depleted than the $z_{low}$ sample,
we expect very minimal evolution in the depletion levels of the
damped systems and no significant impact 
on any of our conclusions. The effects of biasing due to dust obscuration
are more difficult to address.  Note in Figure~\ref{FeHvsNHI} the 
absence of any $\N{HI} > 10^{21} \cm{-2}$ systems with $\feh \sim -1$.
While this may be due to small number statistics or 
that very few regions exist in the early universe with large
$\N{HI}$ and $\feh \gtrsim -1$, the trend could also be explained by dust 
obscuration.
Fall \& Pei (1993) have presented an excellent framework for addressing the
effects of dust depletion on damped \lya statistics. 
Their calculations indicate that if 
the logarithmic scatter in the dust-to-gas ratio $k$ is small
(less than 1 dex), then only a small correction to the mean optical
depth and in turn to $<Z>$ is required (Fall 1999).
For a constant dust-to-metals ratio -- implied by the nearly constant
[Zn/Fe] values -- the logarithmic scatter in $k \approx \sigma(\feh)$
and we have shown $\sigma(\feh) \leq 0.5$ for the two samples.
Therefore, we expect dust obscuration to have a minimal effect
($<0.2$~dex) on our results.

The results in this Letter on the evolution of the metallicity
of neutral gas in the Universe 
and individual protogalaxies present an unexpected picture.
A number of groups have estimated the chemical evolution of neutral gas
at high redshift (\citep{mny96,edmns97,pei99}) and essentially every
treatment predicts a substantial ($> 0.5$~dex) increase in the mean
metallicity from $z = 4$ to $z = 2$. 
While a 0.5~dex evolution is consistent with our results at the 3$\sigma$
level, the current observations favor a very mild evolution in $<Z>$.
If future observations lend further support for this conclusion,
the theoretical models will require significant revision.
Of course, these theoretical treatments
depend sensitively on a number of factors which are uncertain:
(i) the star formation rate, (ii) the IMF, (iii) the mass distribution of
protogalaxies, (iv) the loss of metals to the IGM, (v) the yield of
various elements, etc.
Therefore, there is considerable theoretical freedom to bring the models
into agreement with the observed lack of evolution. Nonetheless,
the Lyman break galaxies offer incontrovertible evidence 
that significant star formation is taking place from $z = 3 - 4$
(\citep{std98})
such that the total metal content of the Universe must be increasing.
Unless these metals are enriching only ionized regions
(an unlikely scenario), then to explain the minimal evolution in $<Z>$
the total HI content of the Universe must be increasing at nearly the same
rate as the metal content. It is intriguing to note 
that this is qualitatively consistent with 
the evolution of $\Omega_{gas}$ observed by
Storrie-Lombardi \& Wolfe (2000) for the damped \lya systems.  

The other statistical moments are sensitive to the chemical
enrichment history within individual galaxies.
Comparing the unweighted mean with the weighted mean
we find that $<\feh>$ is less than $<Z>$ at both epochs.  
While the difference is
not large ($\approx 0.1 - 0.2$~dex), it does highlight the fact that
many of the $\N{HI} < 10^{21} \cm{-2}$ systems exhibit low
metallicity.  In particular, in the $z>3$ sample 
only 2 of 15 damped systems show [Fe/H] $> -1.5$~dex. 
One possible explanation for the difference
is systems which have just formed have preferentially 
low $\N{HI}$ and [Fe/H].
The trend is also suggestive of the correlation Cen \& Ostriker (1999)
find between overdensity and metallicity in their numerical simulations.
The problem remains, however, in explaining why the highest metallicity
systems of the $z_{low}$ sample also have low $\N{HI}$.
Lastly, recall that there is tentative support for an evolution in both
the scatter and $<\feh>$ with the $z_{low}$ sample yielding larger values.  
If more recently formed systems 
tend to have lower metallicity, then the evolution
may easily be explained by a larger mean and scatter in the age
of the damped \lya systems at $z \approx 2$.  Furthermore, the
systems at $z \approx 2$ may have larger masses and a greater variety of
morphologies.   

Finally, we stress that only two systems from the full sample
have [Fe/H]~$< -2.5$ and the large majority show [Fe/H]~$> -2$.  
As first noted by Lu et al.\ (1997), there appears to be a threshold
to the minimum metallicity of the damped \lya systems at $\approx$ 1/100
solar metallicity.
Therefore, even out to $z \approx 4.5$
there is no evidence for damped \lya systems with primordial abundance.
This places a further constraint on chemical evolution models.
As we probe higher and higher redshift without detecting primordial gas,
one may be forced toward one of the following conclusions:
(i) star formation proceeds rapidly
($< 10^7$ years) to bring the metallicity to 1/100 solar
after the formation of a damped system; 
(ii) either the damped system or its progenitors have been undergoing
star formation for a lengthy time; and/or
(iii) a generation of Population III stars has pre-enriched all of the gas.

The future prospects for improving the $z>3$ observational sample are 
excellent.  While further progress with HIRES and similar instruments
is limited by the faintness of $z>4$ quasars, 
the new ESI instrument on Keck~II will be ideal for surveying the $\N{HI}$
and metal content of a large ($N>20$) sample of very high redshift 
damped \lya systems.
We intend to pursue such a program over the next few years, taking full
advantage of the ever increasing sample of known $z>4$ quasars
(\citep{fan99a}).

\acknowledgments

We would like to thank A. McWilliam, E. Gawiser, M. Pettini,
and M. Fall for insightful discussion and comments.
We thank T. Barlow for providing the HIRES reduction package.
We acknowledge the very helpful Keck support staff for their efforts
in performing these observations. J.X.P. acknowledges support from a
Carnegie postdoctoral fellowship.

\end{document}